\documentclass{elsart}
\usepackage{natbib}

\usepackage{epsfig}

\usepackage{amssymb}

\begin{document}

\begin{frontmatter}


% use the thanksref command within \title, \author or \address for footnotes;}

\title{Spectral analysis of 3D MHD models of coronal structures}


\author{Pia Zacharias}
\address{Kiepenheuer Institut f\"ur Sonnenphysik, Sch\"oneckstrasse 6,
  79104 Freiburg, Germany}
%\corauth[cor]{Corresponding author}
%\thanks[footnote2]{Additional information regarding the corresponding author}
\ead{pia@kis.uni-freiburg.de}
%url can be given like this
%\ead[url]{http://authors.elsevier.com/locate/latex}

\author{Sven Bingert and Hardi Peter}
\address{Kiepenheuer Institut f\"ur Sonnenphysik, Sch\"oneckstrasse 6,
  79104 Freiburg, Germany}
%\thanks[footnote3]{Additional information about the second and third authors}
\ead{bingert@kis.uni-freiburg.de, peter@kis.uni-freiburg.de}

%\author{More Authors\thanksref{footnote4}}
%\address{Address of the co-authors}
%\thanks[footnote4]{Additional information about the co-authors}
%\ead{more@email.addresses}

\begin{abstract}
% Text of abstract
We study extreme-ultraviolet emission line spectra derived from three-dimensional magnetohydrodynamic
models of structures in the corona. In order to investigate the effects of increased magnetic activity at
photospheric levels in a numerical experiment, a much higher magnetic flux density is applied at the photosphere as compared to the Sun. 
%One could consider our setup as a ``quiet Sun'' with a level of magnetic flux
%increased by about a factor of four relative to the Sun. 
Thus, we can expect our results to highlight
the differences between the Sun and more active, but still solar-like
stars.
We discuss signatures seen in extreme-ultraviolet emission lines synthesized from these models and
compare them to observed signatures in the spatial distribution and
temporal evolution of Doppler shifts in lines formed in the transition
region and corona. This is of major interest to test the quality of the
underlying magnetohydrodynamic model to heat the corona, i.e. currents in the corona driven by
photospheric motions (flux braiding).

\end{abstract}


\end{frontmatter}

\parindent=0.5 cm

% main text
\section{Introduction}

The mechanism of
field line braiding as a heating mechanism of the solar corona has been
discussed for a long time. However, three-dimensional numerical models of the solar corona and solar-like stars
based on magnetohydrodynamics 
only became available in recent years
(Gudiksen \& Nordlund \citeyear{gudiksen+nordlund:2002}, \citeyear{gudiksen+nordlund:2005b},
\citeyear{gudiksen+nordlund:2005a}). A small active region was described
in a numerical experiment by Gudiksen \& Nordlund (\citeyear{gudiksen+nordlund:2002}) resulting in a
corona consisting of small loop-like structures. In order to investigate
the general appearance of the synthesized corona, synthetic spectra can be
produced from the simulated data (Peter et al. \citeyear{peter+al:2004}, \citeyear{peter+al:2006}) and can then be compared to
spectroscopic observations of the Sun.

In this work, we model the atmosphere above an active region in a three-dimensional box that extends from the photosphere to the corona. The lower boundary magnetic field consists of two parts, i.e. an active region magnetic field and a quiet Sun network magnetic field. The active region magnetic field is the same as in Gudiksen \& Nordlund (\citeyear{gudiksen+nordlund:2002}), i.e. a spatially downscaled magnetogram of Active Region 9114 observed by the Michelson Doppler Imager (MDI; Scherrer et al. \citeyear{scherrer:1995}) onboard the Solar and Heliospheric Observatory (SOHO) on August 8, 2000. The downscaling was performed in order to fit the active region into the computational domain (50x50 Mm$^2$  horizontally). As the downscaling suppresses the small-scale network fields, a typical quiet Sun MDI high-resolution magnetogram was added. In order to highlight the interaction of the active region with the neighboring quiet Sun, the quiet Sun flux density was enhanced by a factor of four. By this a generic magnetogram of an active region is produced that allows us to study the interaction of magnetic structures on small scales. \rm As a lower boundary condition a time-dependent velocity field with properties close to solar granulation (Gudiksen \& Nordlund \citeyear{gudiksen+nordlund:2002}) \rm is imposed. 
Peter et al. (\citeyear{peter+al:2004}) reproduced the observed variation of line shifts with temperature on the Sun from three-dimensional numerical models, especially the temperature of the maximum redshift. This temperature ($\log T \approx $5.3) compares well to the values found for active region and quiet Sun observations (Peter \& Judge \citeyear{peter+judge:1999}, Teriaca et al. \citeyear{teriaca:1999}). Therefore, our numerical experiment can give a first indication of what to expect in a situation with a higher magnetic complexity.

Active stars with convective envelopes show a complicated surface magnetic field distribution. High-resolution spectra are needed to separate the measured magnetic flux $Bf$, into the magnetic field density $B$ and the magnetic filling factor $f$. Saar (\citeyear{saar:1996}) investigated the behavior of stellar surface magnetic fields of main sequence dwarf stars and found a correlation between the measured magnetic flux $Bf$, and the rotation period of a star. The stellar magnetic flux $Bf$, as well as the filling factor $f$ of stars that rotate slower than the saturation threshold grows with larger rotation rate. The fluxes reach values up to 100 times larger than those observed for the Sun and imply an increased heating of the upper layers of the stars' atmospheres. Saar (\citeyear{saar:1996}) claims that saturation occurs in the filling factor $f$, but not in the magnetic flux $Bf$. From this result one could conclude that a star rotating at the saturation threshold is completely covered with magnetic field ($f$=1) and exhibits a much higher magnetic complexity compared to the Sun, where the magnetic flux is concentrated in relatively small regions of strong fields. Doppler imaging and Zeeman Doppler imaging give evidence for starspots and stellar surface magnetic flux at high latitudes (Donati et al. \citeyear{donati+al:1999}), but also for spots and magnetic fields at low latitudes (e.g. Barnes et al. \citeyear{barnes+al:1998}) suggesting more complex surface magnetic field topologies than found on the Sun.
The setup of the model we use in this study to investigate properties of the synthesized emission lines can be considered as a step towards higher magnetic complexity and higher magnetic flux at surface levels, and thus as a step towards investigating the coronae of stars more active than our Sun.
\rm

The temperatures and
densities in the simulation box as well as the velocity components and the magnetic field components are
accessible at each grid point for every time-step of the simulation. The
model temperatures and densities cover several orders of magnitude. The
well-known rise in temperature and the drop in density from the photosphere
to the corona are reproduced well in the box. 
We synthesize emission-line spectra from such coronal models
and investigate average properties of these spectra, such as intensities and
Doppler shifts.\\

\section{MHD model and spectral analysis}
In this model the heating of the corona is based on the braiding of
magnetic field lines due to motions in the photosphere, as first suggested
by \citet{parker:1972}. The braided magnetic field produces currents that
are dissipated by Ohmic heating.
{We are using the Pencil Code (Brandenburg \& Dobler \citeyear{brandenburg+dobler:2002}}), a high-order finite difference code (sixth order in space and third order in time) for solving the MHD equations in a compressible medium. The three-dimensional magnetohydrodynamic model accounts properly for the energy
balance, especially for heat conduction and radiative losses, allowing us to reliably synthesize the profiles of extreme-ultraviolet emission lines observable
with current EUV spectrometers, i.e. the Solar Ultraviolet Measurement of Emitted Radiation (SUMER) instrument onboard SOHO (Wilhelm et al. \citeyear{wilhelm:1995}) and the Extreme-ultraviolet Imaging Spectrometer (EIS) onboard Hinode (Culhane et al. \citeyear{culhane+al:2007}).

The computational box extends 50 Mm x 50 Mm in the horizontal
direction and 30 Mm in the vertical direction covering the atmosphere from
the photosphere to the corona with a 256$^3$ grid that is non-equidistant
in the vertical direction. The vertical grid spacing goes down to 80 km in
the low transition region ($\log T \approx 4.3$) around the maximum temperature gradient with a spacing in temperature of 0.18 dex on a log $T$ scale. For the calculation of the emission lines, the vertical
spacing of the grid is not sufficient. Therefore, a spatial interpolation \citep{peter+al:2006} has to be applied.

We determine the emissivities, intensities and Doppler velocities
for a number of extreme-ultraviolet emission lines (Table 1) from the
temperatures and densities at each point of the interpolated grid in the box. Also shown in Table 1 are the line formation temperatures of the respective line, i.e. the temperature where the maximum of the contribution function is found. This is roughly comparable, but not identical to the temperature of the maximum ionization fraction. The lines have been selected for two reasons. First, they cover the range of temperatures in the transition region, i.e. from $T\approx 10^4-10^6$ K. Second, the lines are observable with spectrometers such as SUMER/SOHO and EIS/Hinode, and are major lines in the bandpasses of the future Atmospheric Imaging Assembly (AIA) for the Solar Dynamics Observatory (SDO). \rm The lines
are optically thin, and predominantly excited by electron collisions.  
The emissivity of the transition from an upper level 2 to a lower level 1
is given by $\epsilon=h\nu n_2 A_{21}$, where $h\nu$ is the energy, $n_2$ is
the number density of the upper level, and $A_{21}$ is the Einstein
coefficient. To calculate the line emissivities from the MHD model we employ the CHIANTI package (Dere et al. \citeyear{dere+al:1997}, Landi et al. \citeyear{landi+al:2006}). \rm For the calculation of the number density of the upper level
ionization equilibrium is assumed and the Mazzotta et al. (\citeyear{mazzotta+al:1998}) ionization fractions are used. The validity of ionization equilibrium for modeling synthetic spectra from three-dimensional MHD models is a critical point and has been investigated by Peter et al. (\citeyear{peter+al:2006}). Their discussion of ionization and recombination times in comparison to the dynamic timescale, i.e. the time to cross the line formation temperature, showed that the ionization and recombination times in the coronal and transition region plasma are mostly smaller than the typical hydrodynamic timescales, indicating that the approximation of ionization equilibrium is not too bad. However, for future models the inclusion of non-equilbrium ionization is very important. Since we do not consider line ratios here, the results do not depend on the specific abundances used in this study. We refer to the photospheric elemental abundances (Grevesse \& Sauval \citeyear{grevesse+sauval:1998}) \rm in CHIANTI. \rm
The spectral profiles are assumed to be Gaussian with a line width
corresponding to the thermal width $(2 k_B T/m)^{1/2}$, where $T$ is the
temperature at each grid point and $m$ is the mass of the ion emitting the
line. The spectra are integrated along a given line of sight to obtain
two-dimensional spectral maps. The profile moments with respect to velocity
are evaluated in order to obtain maps of line intensity, line shift, and
line width. Based on a 30 minute time series of maps of line intensities and line shifts, the root-mean-square (rms) fluctuations
at each spatial pixel within the respective map are evaluated. We investigate
the correlation between the average Doppler shift and relative brightness
variabilities of the emission lines as previously done by \citet{brkovic+al:2003}, who studied time series of spectra and images of quiet-Sun regions observed using the SUMER spectrometer and the Coronal Diagnostic Spectrometer (CDS) (Harisson et al. \citeyear{harrison+al:1995}) onboard the SOHO spacecraft.\rm

\section{Results}
Figure 1 shows the emissivities of C\,{\sc{iv}} (1548 \AA) and Mg\,{\sc{x}}
(625 \AA) in the box integrated along
the x- and y-direction 30 minutes after the start of the simulation. This would be equivalent to an observation at the limb. The emissivities 
are plotted on a logarithmic scale according to 0.1-10 times the average
intensity (white: emission, black: no emission). Figure 2 shows the
emissivities of C\,{\sc{iv}} (1548 \AA) and Mg\,{\sc{x}}
(625 \AA) integrated along the z-direction, i.e. when looking at the box
from straight above, and the spatial maps in
Doppler shift after 30 minutes of simulation time. This corresponds to an observation near disk center. The Doppler maps are
scaled from -20 km s$^{-1}$ (red; downflows) to +20 km s$^{-1}$ (blue; upflows).

\begin{figure}
\begin{center}
\includegraphics[width=14.5cm]{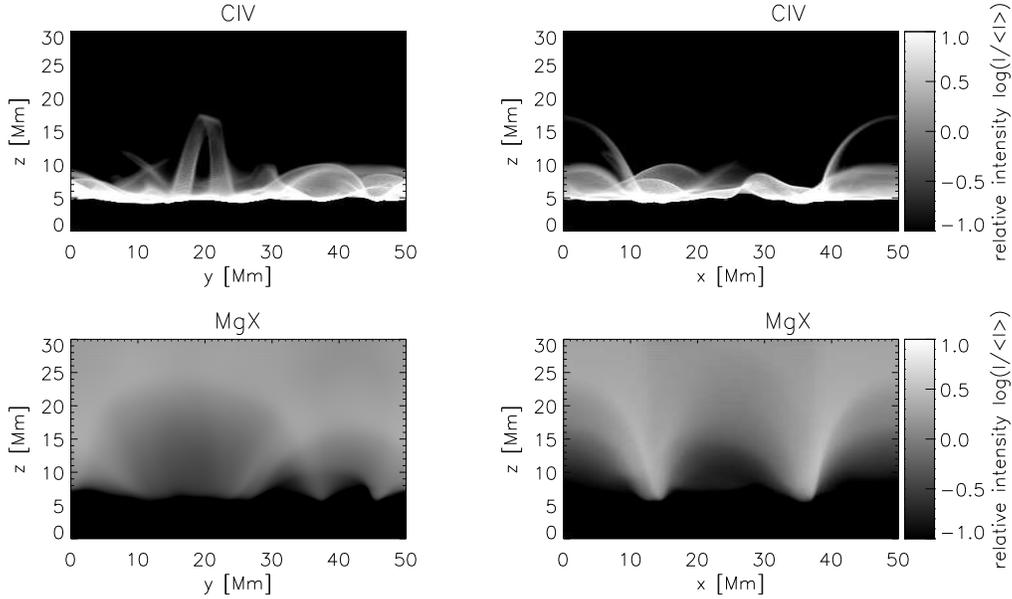}
\end{center}
\caption{Maps in line intensity of the transition region line C\,{\sc{iv}}
  (1548 \AA) formed at $T \approx 10^5$ K and the coronal line Mg\,{\sc{x}} (625
  \AA) formed at $T \approx 10^6$ K. The panels show the intensity images when
  viewing the box from the sides, i.e., along the x and y axis.\label{figure1}}
\end{figure}

\begin{figure}
\begin{center}
\includegraphics[width=14.5cm]{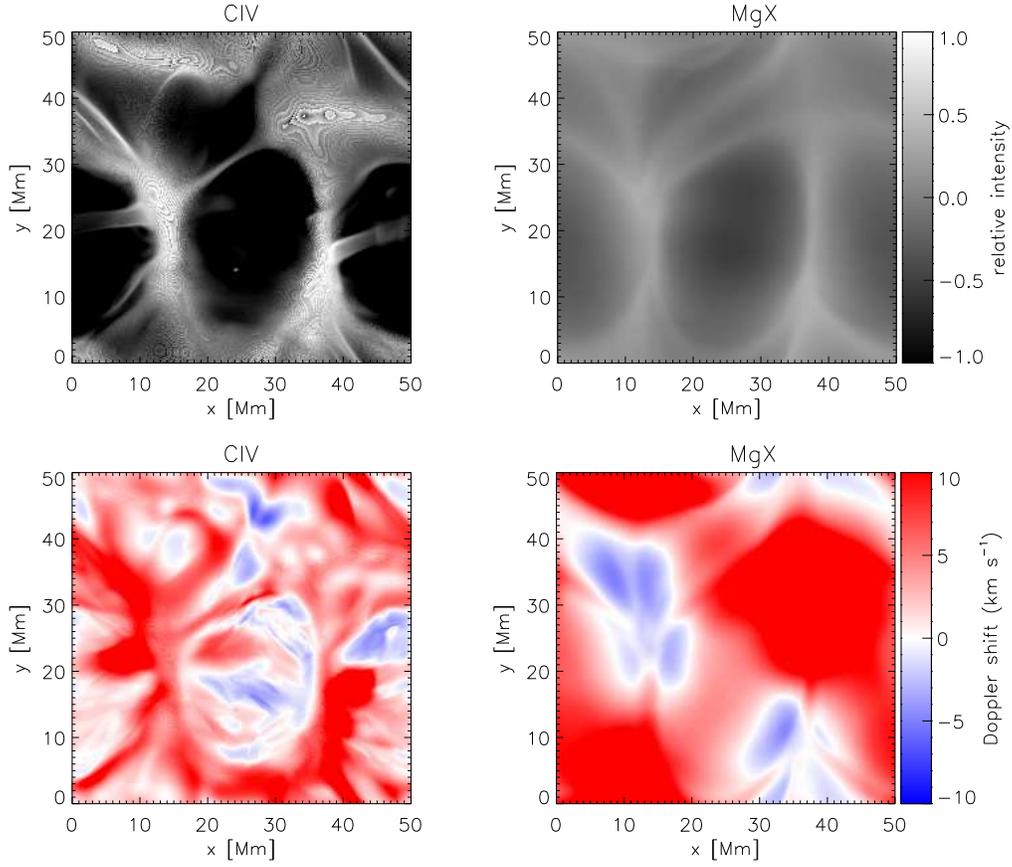}
\end{center}
\caption{Maps of line intensity and Doppler shift of C\,{\sc{iv}} (1548 \AA) and Mg\,{\sc{x}} (625 \AA) when looking at the box from straight above.\label{figure2}}
\end{figure}

\begin{figure}
\begin{center}
\includegraphics[width=14.0cm]{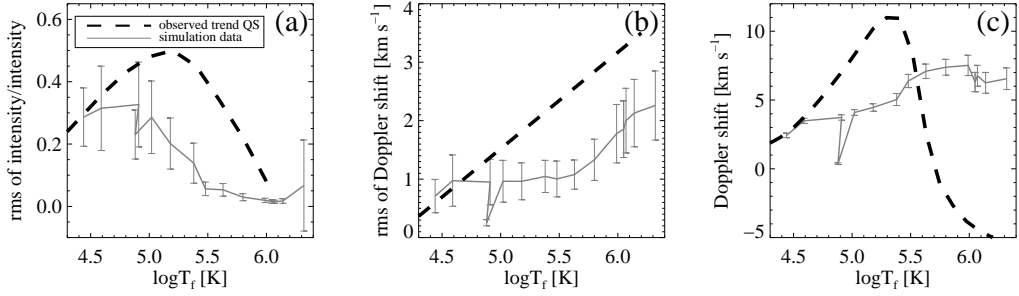}
\end{center}
\caption{rms fluctuations of intensity (a) and Doppler shift (b) as a function of line formation temperature for a 30 minute
  time-series of spectra (solid lines: simulation data, dashed lines: observational results). The error bars in (a) and (b) represent the root-mean-square deviation of the spatially averaged rms intensities and rms Doppler shifts. \rm (c) shows the Doppler shifts for each line averaged over time. The error bars indicate the standard deviation of the time-averaged Doppler shifts. \label{figure3}} 
\end{figure}

Figure 3 displays the spatially averaged values of the rms intensity fluctuations (panel a) and the the spatially averaged values of the
root-mean-square (rms) Doppler shifts (panel b) of the 30 minute time series as a function of line formation temperature for
the extreme-ultraviolet emission lines from Table 1.
Also shown are the time-averaged Doppler shifts as a function of line
formation temperature of the emission lines (panel c).% as previously studied by \citet{brkovic+al:2003}. 

The signatures of our model corona (solid lines) with a higher magnetic flux density at photospheric levels than the Sun 
differ significantly from the signatures on the quiet Sun (dashed lines; 
observed data in Fig 3a and 3b from Brkovic et al. \citeyear{brkovic+al:2003}, observed data in Fig. 3c from Peter \& Judge \citeyear{peter+judge:1999}).
The quiet Sun rms intensity fluctuations show a clear maximum around $3 \times 10^5$ K (Fig. 3a). 
The numerical experiment with a higher photospheric magnetic flux density 
than on the quiet Sun shows fluctuations that are less strong at higher temperatures and a maximum at lower temperatures than the observed solar data. 
The rms Doppler shift fluctuation
as a function of line formation temperature (Fig. 3b) shows 
a monotonic increase for both, the simulated data (solid line) and the observed data (dashed line). However, the increase of the rms
Doppler shift for the simulated data is less strong than for the observed data on the quiet Sun.
The time-averaged Doppler shifts of the simulation increase with line formation temperature up to 7 km s$^{-1}$ at log$T \approx 10^6$ K, whereas on the quiet Sun, a steady increase of line redshift up to 10 km s$^{-1}$ at log$T$=5.2 is observed (Fig. 3c). We take this as an indication \rm that stars with higher activity levels than quiet Sun regions show a maximum redshift at higher temperatures as compared to the quiet Sun. However, lacking an observed Sun-as-a-star spectrum one has to act with caution when comparing absolute line shifts of the Sun and solar-like stars. According to Peter (\citeyear{peter:2006}), the observed disk-center redshifts for the Sun exhibit a net line shift in the middle transition region three times higher as compared to full-disk values. Peter's (2006) study also revealed that the net line shift of solar transition regions lines is at least a factor of four lower than the net shifts found on the solar-like star $\alpha$ Cen A, probably indicating a higher magnetic activity in the chromospheric network of $\alpha$ Cen A.\\ 
%Existing stellar observations are not conclusive on this issue.\\
The trend of redshift with line formation temperature of late-type (dwarf) stars has been investigated in many works (e.g. Wood et al. \citeyear{wood+al:1997}, Redfield et al. \citeyear{redfield+al:2002}, Sim \& Jordan \citeyear{sim+jordan:2003}).
Wood et al. (\citeyear{wood+al:1997}) found that emission line redshifts of $\alpha$ Cen A and B increase with line formation temperature up to log$T$=5.2 almost indentical to those observed for the Sun and Procyon. Above log$T$=5.2, the redshifts decrease for the Sun, $\alpha$ Cen A, and $\alpha$ Cen B. However, for Procyon, they continue to increase. 
Redfield et al. (\citeyear{redfield+al:2002}) investigated O\,{\sc{vi}} lines formed at log$T$=5.5 for seven late-type dwarf stars including Procyon, AB Dor, $\alpha$ Cen A, $\alpha$ Cen B, $\epsilon$ Eri, and AU Mic. Most stars show an increase in velocity offset with increasing line formation temperature and sooner or later a decrease with increasing temperature, the same general behavior as for solar line shifts (Peter \& Judge. \citeyear{peter+judge:1999}).  
%and found increasing redshifts with line formation temperature for Procyon, AB Dor, and $\alpha$ Cen A. 
However, the O\,{\sc{vi}} line shift in $\alpha$ Cen A and $\epsilon$ Eri is highly redshifted with respect to other high-temperature lines, such as O\,{\sc{v}}. For the cooler stars in their sample the O\,{\sc{vi}} redshifts are found to be close to zero. At least for $\alpha$ Cen A this is a result in contradiction to the one of Wood et al. (\citeyear{wood+al:1997}).
In Redfield et al. (\citeyear{redfield+al:2002}) transition region lines of stars like $\alpha$ Cen A and Procyon do not suggest a drop-off in redshift over the observed temperature range (up to log$T$=5.5; Redfield et al. \citeyear{redfield+al:2002}) as found for solar data (maximum redshift around $\log T$=5.2), a trend consistent with the model results shown in Fig 3. As we do not explore a range of magnetic fluxes at the stellar surface, we cannot comment on the connection of e.g. magnetic activity and temperature or emission measure in stellar coronae yet. 

\begin{table}[h]
\caption{Wavelengths and line 
  formation temperatures of emission lines synthesized in this study \label{Tab}.}
\begin{center}
\begin{tabular}{lrr}
\hline
\hline
line            &  wavelength [\AA] & $\log(T_f[\rm K])$ \\
\hline
Si\,{\sc{ii}}   & 1533.4            & 4.44\\
Si\,{\sc{iv}}   & 1393.8            & 4.88\\
C\,{\sc{ii}}    & 1334.5            & 4.59\\
C\,{\sc{iii}}   &  977.0            & 4.91\\
C\,{\sc{iv}}    & 1548.2            & 5.02\\
O\,{\sc{iv}}    & 1401.2            & 5.18\\
O\,{\sc{v}}     &  629.7            & 5.38\\
O\,{\sc{vi}}    & 1031.9            & 5.48\\
Ne\,{\sc{viii}} &  770.4            & 5.80\\
Mg\,{\sc{x}}    &  624.9            & 6.05\\
Mg\,{\sc{x}}    &  609.8            & 6.05\\
Fe\,{\sc{viii}} &  185.1            & 5.63\\
Fe\,{\sc{x}}    &  184.5            & 6.99\\
Fe\,{\sc{xi}}   &  188.2            & 6.07\\
Fe\,{\sc{xii}}  &  195.1            & 6.14\\
Fe\,{\sc{xv}}   &  284.2            & 6.32\\
\hline
\hline
\end{tabular}
\end{center}
\end{table}

\section{Conclusions}
We present results of extreme-ultraviolet emission line spectra synthesized
from three-dimensional 
magnetohydrodynamic models of the corona with a photospheric magnetic flux density consisting of a quiet Sun part with a magnetic flux density four times larger than on the quiet Sun and an active region part which involves a spatially downscaled magnetogram. We show average properties 
of the synthetic spectra and compare them to observed properties of solar line profiles.

% from SUMER/SOHO.
By investigating maps of intensity and Doppler shift when integrating
along the line of sight through the box, one can study the morphology
of the transition region and corona. Here, we show images of the
transition region line C\,{\sc{iv}} (1548 \AA) formed at $T\approx 10^5$
K and the coronal line Mg\,{\sc{x}} (625 \AA) formed at $T\approx 10^6$ K.
The coronal emission appears diffuse and continuous in the upper part
of the box, whereas one finds cool low-lying loops concentrated in the transition
region.
This result matches observation of the quiet Sun, where the
transition region emission is confined to the chromospheric magnetic
network, while the emission from the hot corona appears diffuse.
The dopplergram of the transition region line is found to be highly-structured with a
much smaller amplitude in redshift than in the corona. These findings
agree well with solar observations. They support the idea that the coronal heating process is due to the braiding of magnetic field lines and results from photospheric footpoint motions. The Doppler shifts follow from the driving of the corona by the footpoint motion of the magnetic field lines that has been implemented in the magnetohydrodynamic model. As an effect of increased magnetic activity in the numerical simulation, the maximum redshift is shifted towards higher temperatures. Peter et al. (\citeyear{peter+al:2004}) showed earlier that three-dimensional numerical simulations can reproduce the observed variation of line shifts with temperature on the Sun. Our numerical experiment with increased magnetic complexity indicates that stars with higher levels of activity might show maximum redshifts at higher temperatures. Unfortunately, observational results on redshifts of stars at various levels of activity are not conclusive yet.

\bibliography{asr_literature}
\bibliographystyle{elsart-harv}

\clearpage

\end{document}